\documentclass[runningheads,a4paper]{llncs}

\usepackage{amssymb}
\usepackage{graphicx}

\usepackage[utf8]{inputenc} 
\usepackage{amsmath}
\usepackage{eqnarray}
\usepackage{bm}
\usepackage{mathtools}
\usepackage{xcolor}
\usepackage{xfrac}
\usepackage{esint}
\usepackage{mathcomp}
\usepackage[rightcaption]{sidecap}
\usepackage[margin=2cm]{caption}  		
\usepackage[hidelinks]{hyperref}
\usepackage{afterpage}
\usepackage{tikz}
\usepackage{fancyhdr}
\usepackage{acro} 
\usepackage[lined,ruled,vlined,linesnumbered,spanish, onelanguage]{algorithm2e}
\usepackage{listings}
\usepackage{float}
\usepackage{multirow}
\usepackage{tabularray}
\usepackage[T1]{fontenc}

\usepackage{url}
\urldef{\mailsa}\path|arojas@campus.ungs.edu.ar|    
\urldef{\mailsb}\path|brendaamareco@gmail.com|    

\newcommand{\keywords}[1]{\par\addvspace\baselineskip
\noindent\keywordname\enspace\ignorespaces#1}

\begin{document}

\mainmatter  

\title{Shannon Entropy\\ is better Feature than Category and Sentiment\\ in User Feedback Processing}



%
%
\author{Andr\'es Rojas Paredes%
\thanks{Work supported by project UNGS 30/1147} 
\and Brenda Mareco}
%

\institute{Instituto de Ciencias, Universidad Nacional de General Sarmiento,\\
Juan María Gutiérrez 1150, CP1613, Los Polvorines, Argentina\\
\mailsa\\
\mailsb\\
\url{http://www.ungs.edu.ar/ici}}

%
%


\pagestyle{empty}
 
\maketitle

\begin{abstract}
App reviews in mobile app stores contain useful information which is used to improve applications and promote software evolution. This information is processed by automatic tools which prioritize reviews. In order to carry out this prioritization, reviews are decomposed into features like category and sentiment. Then, a weighted function assigns a weight to each feature and a review ranking is calculated. Unfortunately, in order to extract category and sentiment from reviews, its is required at least a classifier trained in an annotated corpus. Therefore this task is computational demanding. Thus, in this work, we propose Shannon Entropy as a simple feature which can replace standard features. Our results show that a Shannon Entropy based ranking is better than a standard ranking according to the NDCG metric. This result is promising even if we require fairness by means of algorithmic bias. Finally, we highlight a computational limit which appears in the search of the best ranking.

\end{abstract}

\section{Introduction}

In this work we study user feedback processing in requirements engineering. Previous work has found that user feedback contains information that is useful to analysts and app designers, such as user requirements, bug reports and ethical concerns (see e.g. \cite{ethical}). In order to extract useful information from reviews, and due to the unstructured nature and the large amount of available reviews, automated tools are required to classify reviews according to their importance (see e.g. \cite{ALERTme} and \cite{odyssey}). 

A user feedback processing pipeline receives raw app reviews as input and returns a review ranking as output, and has four main stages:
\begin{enumerate}
    \item \textbf{\emph{Preprocessing}} and \textbf{\emph{Feature Extraction}}: prepares raw reviews for next stages, main features are calculated in this stage.
   
    \item \emph{\textbf{Ranking}}: prioritize reviews according to a weighted function (see \cite{ALERTme}). In this work, we focus on weights which are assigned to features like category and sentiment. A weight represents the importance of a feature against other feature used in the weighted function.
        
    \item \emph{\textbf{Quality Testing}}: evaluates the quality of a review ranking using the NDCG -- Normalized Discounted Cumulative Gain metric (see \cite{ndcg}). NDCG is a score between 0 and 1, and it is defined as follows:

\begin{equation}
NDCG=\frac{DCG}{IDCG} \text{\textcolor{white}{espacio}} DCG= \displaystyle\sum_{i=1}^{n} \frac{rank_i}{log_2(i)}
\end{equation}
where $rank_i$ is the calculated ranking for review $i$. The value $IDCG$ is the ideal DCG according to a manual annotated ranking. Thus, we compare an algorithmic ranking against a manual ranking by means of the NDCG metric.
    
    On the other hand, since fairness is an important concern in app reviews (see \cite{fairness2} and \cite{gender}), this stage of the pipeline also detects algorithmic bias in the produced ranking.
\end{enumerate}

\subsection{Research Questions} %
\label{sec:objetivos-preguntas-investigacion}

The main objective of this work is to calculate the best weights for a weighted function in order to maximize the NDCG metric against a set of reviews which are manually ranked by experts. Thus, according to the stated objective this work answers the following research questions:
\begin{itemize} 
    \item \label{R.Q.1} \textbf{R.Q.1:} What is the best combination of weights for the standard features used in a weighted function (Category, Sentiment, Rating and Length) such that the NDCG metric is maximized?
    \item \label{R.Q.2} \textbf{R.Q.2:} Is Shannon Entropy a feature which could replace standard features like Sentiment and Category?
    \item \label{R.Q.3} \textbf{R.Q.3:} Is there any computational limit if we increase the number of digits precision in the weights?
    \item \label{R.Q.4} \textbf{R.Q.4:} How algorithmic bias and bias mitigation affect the quality of the ranking?
\end{itemize}

\section{Methodology} \label{Capitulo_3}


\subsection{Data exploration and preparation}

\subsubsection{Data collection and annotation process}

We use the dataset from \cite{userFeedbackAppStore} and \cite{masterProjectUserFeedback}. This dataset contains reviews from Apple App Store users from eight countries (Australia, Canada, Hong Kong, India, Singapore, South Africa, the United Kingdom and the United States), on seven applications, with a total of 59203 reviews between May 1 and June 30, 2017.

The dataset includes 160 randomly selected reviews (20 per country), manually ranked by an annotator on a scale from 0 to 3.

\subsection{Experiment design}
In order to answer our research questions in Section \ref{sec:objetivos-preguntas-investigacion} four experiments were carried out. Section \ref{Results} below shows our min results.

\subsection{Pipeline implementation}
\label{pipelImplementationSection}

A pipeline was implemented in order to execute our experiments. Figure \ref{fig:Pipeline} below shows a C\&C (Components and Connectors) architecture of our pipeline.

\begin{figure}[H]
	\centering
	\includegraphics[scale=0.35]{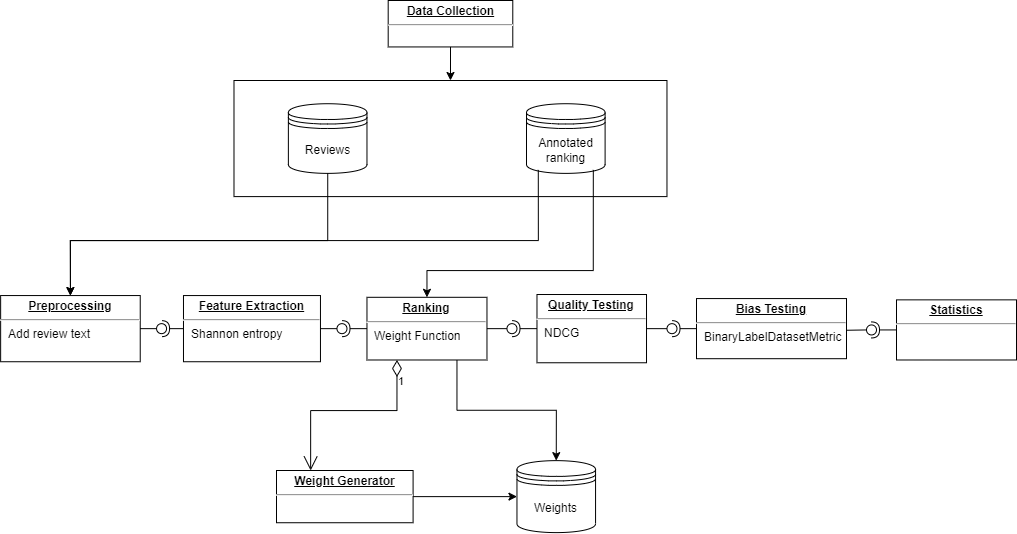}
	\caption{User Feedback Processing Pipeline}
 \label{fig:Pipeline}
\end{figure}

\subsubsection{Preprocessing}. This component adds a new column to the \emph{Annotated ranking} dataset with the full text of the reviews from the \emph{Reviews} dataset.

\subsubsection{Feature extraction}.
\label{sec-extraccionAtributos}
This component calculates the Shannon Entropy from the review text and adds the result as a new column in \emph{Annotated ranking}.

\subsubsection{Ranking}.
\label{subsec:ranking}
This component calculates the ranking of reviews using a weighted function adapted from \cite{ALERTme}. In this work we use the followig features:

\textit{\textbf{Category}}: A review is a (\emph{Bug Report}, \emph{Feature Request} or \emph{Other}).

\textit{\textbf{Sentiment}}: A review sentiment is an integer value in the range $(-2, 2)$, where $2$ is very positive and $-2$ is very negative, and $0$ is neutral.

\textit{\textbf{Score}}: This feature is the number of stars rating an app ($1$ to $5$).

\textit{\textbf{Revlen}}: This feature is the text length.

\textit{\textbf{Entropy}}: This feature is the amount of useful information in a review, calculated with Shannon Entropy formula.

Thus, taking into account these features, our formula to calculate the ranking $R$ of a review $c$ is:

\begin{equation} \label{eq:funcionPesos}
    R(c) =\displaystyle\sum\limits_{i=1}^4 w_{i}*f_{i}(c)
\end{equation}

\label{table:factorRanking}
where $w_{i}$ represents the weight of feature $i$, $f_{i}$ represents the ranking factor of feature $i$ (see Table \ref{table:factorRanking}). Notice that four features of the five defined above are being used in the formula, since initially the calculation was carried out using all the features except \emph{Entropy} and after that the ranking was recalculated by replacing the feature \emph{RevLen} with \emph{Entropy}.

\begin{table}[H]
\centering
\caption{Ranking factor according to each feature}
\resizebox{\linewidth}{!}{%
\begin{tblr}{
  row{1} = {c},
  cell{7}{1} = {c=2}{},
  cell{8}{1} = {c=2}{},
  cell{9}{1} = {c=2}{},
  cell{10}{1} = {c=2}{},
  hline{1-2,7} = {-}{},
}
Feature                                                                                                                                          & ranking factor \\
Category~                                                                                                                                         & 
$
    f_{1} =
    \begin{cases}
     & 1 \text{ if bug report }\\ 
     & \text{0.5 if feature request } \\ 
     & 0 \text{ otherwise } 
    \end{cases}
$
                          \\
Sentiment~                                                                                                                                        & $f_{2}=\frac{1}{\text{sentiment}+3^{[1]}}$                           \\
Score~                                                                                                                                            & 
$f_{3} = \frac{1}{\text{number of stars}}$ \\
Revlen~                                                                                                                                           & $f_{4}=\frac{\text{review length}}{\text{maximum length}^{[2]}}$                           \\
Entropy~                                                                                                                                          & $f_{4}^{[3]}=\frac{\text{review entropy}}{\text{maximum entropy}^{[4]}}$                           \\
\footnotesize{{[}1] 3 is added to the sentiment value in order to avoid division by 0}                                     &                               \\
\footnotesize{{[}2] Refers to the maximum length of all reviews}                                                &                               \\
\footnotesize{{[}3] $f_4$ is repeated in two features because you must choose between Revlen or Entropy} &                               \\
\footnotesize{{[}4] Refers to the highest entropy of all reviews}                                         &                               
\end{tblr}
}
\label{table:factorRanking}
\end{table}

In this work we obtain the combination of weights that maximizes the NDCG metric. An exhaustive search of all possible weight combinations was carried out, calculating and evaluating the ranking for each combination.

The different weight combinations was carried out by a \emph{Weight Generator} component, using a Backtracking algorithm. 






\subsubsection{Quality Testing}
This stage of our pipeline executes a \emph{Quality Testing} component, which calculates the quality of all the rankings previously obtained using the NDCG metric. This metric returns a real value between $0$ and $1$, where values close to $1$ indicate greater similarity with the ranking annotated by experts.

\subsubsection{Bias testing}
In this stage a \emph{Bias Testing} component verifies, using the AIF 360 tool, whether any country is favored by the ranking. If bias is detected, it is mitigated using the Reweighing algorithm (see \cite{reweighing}), and a new ranking is generated.

\subsubsection{Statistics}
\begin{figure}[H]
    \centering
    \includegraphics[width=0.80\linewidth]{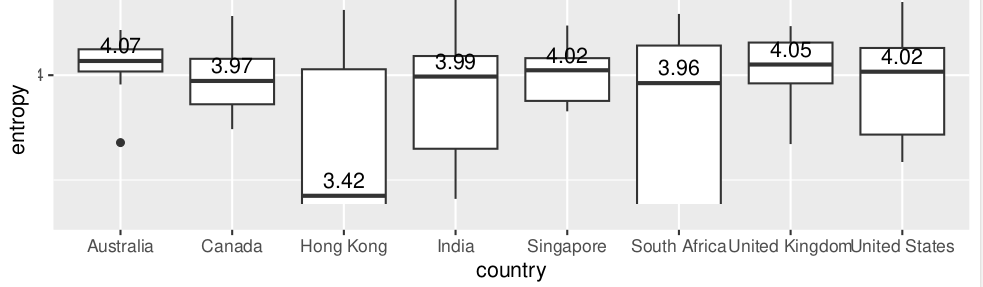}
    \caption{Entropy by country}
    \label{fig:entropy-country}
\end{figure}

In this last stage a \emph{Statistics} component creates plots of the ranking and other plots such as entropy by country (see Figure \ref{fig:entropy-country}).

\section{Results}
\label{Results}

\paragraph{\textbf{Experiment 1: looking for the best ranking using standard features.}}
The best combination of weights we found using two decimal digits precision is
\{$Category$ = 0.7, $Sentiment$ = 0.05, $Score$ = 0.03, $RevLen$ = 0.22 \}, where \emph{Category} is the feature with the most important weight compared to the others features. The NDCG of this weight combination is 0.9790262355941447.

\paragraph{\textbf{Experiment 2: looking for the best ranking with Entropy instead of Length.}}
Using two decimal digits precision (0.01), the weight combination \{$Category$ = 0.27, $Sentiment$ = 0.18, $Score$ = 0.03, $Entropy$ = 0.52 \} has the highest NDCG (0.9816799804069377).

\paragraph{\textbf{Computational limit.}}
In our experiments, a limitation was found when using three decimal digits precision, since each weight can take a value from the set \{0, 0.001, 0.002, \ldots, 0.999, 1.0\} with 1001 possible values. By assigning a weight to each of the four features, there are $1001^{4} = 1004006004001$ possible combinations. For each solution, the ranking must be calculated with the weighted function and evaluated the NDCG, which requires considerably more time and resources by increasing the number of digits precision.
Table \ref{limite-computacional} shows how the pipeline execution time an space increases as the digits precision weight also increase. Table \ref{limite-computacional} below shows the numbers of this computational limit. 

\begin{table}[H]
\centering
\caption{Pipeline execution time and space}
\resizebox{\linewidth}{!}{%
\begin{tblr}{
  cell{6}{1} = {c=4}{},
  cell{7}{1} = {c=4}{},
  hline{1-2,6} = {-}{},
}
Digits Precision &  Combinations   & Solutions & Pipeline execution time & Used disk space \\
0 (1.0)          & $2^{4}$ = 16          & 4          & 0m2.326s  &48.1 kB          \\
1 (0.1)          & $11^{4}$ = 14641      & 286        & 0m2.597s    &841.9~kB        \\
2 (0.01)         & $101^{4}$ =~104060401 & 176851     & 7m44.967s    & 538.0~MB\\    
3 (0.001)         & $1001^{4}$ =~1004006004001 & 1706311567 (aprox.)$^{[1]}$     & 52 days (aprox.)$^{[2]}$ & 5069.12 GB (aprox.)\\
\footnotesize{{[}1] It was obtained by calculating $\frac{1001^{4}\cdot 176851}{101^{4}}$, where 176851 is the number of solutions in $101^{4}$ combinations} \\
\footnotesize{{[}2] It was obtained by calculating $\frac{1001^{4}\cdot 4486459.6}{101^{4}}$, where $4486459.6$ is the equivalent $7m44.967s$ in seconds} \\
\end{tblr}
}
\label{limite-computacional}
\end{table}

\paragraph{\textbf{Experiment 3: Country Bias Detection in our ranking.}}
Country bias was detected in the ranking annotated by experts and in the best ranking calculated with the weighted function, using the set of features \{\emph{Category}, \emph{Sentiment}, \emph{Score}, \emph{Entropy}\} and the weights \{Category=0.27, Sentiment=0.18, Score=0.03, Entropy=0.52\}.
Table \ref{Detección de bias del ranking calculado con función de pesos} below shows a bias favoring Australia and Singapore in the best ranking, and, on the other hand, Table \ref{Detección de bias del ranking anotado por expertos table2} shows a bias favoring Hong Kong, India and Singapore in the ranking annotated by experts.

\begin{table}
\centering
\caption{Bias in our ranking}
\resizebox{\linewidth}{!}{%
\begin{tblr}{
  column{3} = {c},
  column{5} = {c},
  cell{1}{4} = {c},
  hline{1-2,4} = {-}{},
}
Privileged country  & Disparate impact           & Bias-d        & Statistical parity          & Bias--s        \\
\textbf{Australia} & \textbf{0.692307692307692} & \textbf{True} & \textbf{-0.2}               & \textbf{True} \\
\textbf{Singapore} & \textbf{0.761904761904762} & \textbf{True} & \textbf{-0.142857142857143} & \textbf{True} 
\end{tblr}
}
\label{Detección de bias del ranking calculado con función de pesos}
\end{table}

\begin{table}
\centering
\caption{Bias in the annotated ranking}
\resizebox{\linewidth}{!}{%
\begin{tblr}{
  row{1} = {c},
  cell{2}{3} = {c},
  cell{2}{5} = {c},
  cell{3}{3} = {c},
  cell{3}{5} = {c},
  cell{4}{3} = {c},
  cell{4}{5} = {c},
  hline{1-2,5} = {-}{},
}
Country privileged  & Disparate impact           & Bias--d        & Statistical parity          & Bias--s        \\
\textbf{Hong Kong} & \textbf{0.746031746031746} & \textbf{True} & \textbf{-0.114285714285714} & \textbf{True} \\
\textbf{India}     & \textbf{0.657142857142857} & \textbf{True} & \textbf{-0.171428571428571} & \textbf{True} \\
\textbf{Singapore} & \textbf{0.584415584415584} & \textbf{True} & \textbf{-0.228571428571429} & \textbf{True} 
\end{tblr}
}
\label{Detección de bias del ranking anotado por expertos table2}
\end{table}


Notice that users from India and Hong Kong wrote more reviews about \emph{Bugs} than other countries. Since this category is the one with the highest weight, and experts considered \emph{Category} as the most important feature when annotating the ranking, it follows a bias in favor of Hong Kong and India according to Table \ref{Detección de bias del ranking anotado por expertos table2}.

In our ranking with a weighted function, bias was detected in Australia, where entropy is the most important feature, explaining the bias in favor of Australia, the country with the highest entropy (see Figure \ref{fig:entropy-country}).


\paragraph{\textbf{Experiment 4: Bias mitigation and how it affects the NDCG value.}}
In this experiment, we start from the ranking obtained with the weighted function that uses the features \{Category, Sentiment, Score, Entropy\} and the weights \{Category=0.27, Sentiment=0.18, Score=0.03, Entropy=0.52\}. In order to mitigate the cases where there was bias (Australia and Singapore) the reweighing mitigation algorithm was applied.

After applying reweighing to the biased rankings, the bias was mitigated. Other rankings were not mitigated because in previous experiments it was observed that mitigating a ranking that has no bias reduces its NDCG (see Table \ref{Resultados después de aplicar mitigación}). Thus, after applying bias mitigation, the NDCG value decreased with respect to those obtained in previous experiments, using the best combination of weights with the features without \emph{Entropy} ($0.9790262355941447$) and with \emph{Entropy} ($0.9816799804069377$). Table \ref{Comparación entre el valor de NDCG} shows these results.

\begin{table}[H]
\centering
\caption{Results after bias mitigation}
\resizebox{\linewidth}{!}{%
\begin{tblr}{
  row{1} = {c},
  cell{2}{3} = {c},
  cell{2}{5} = {c},
  cell{3}{3} = {c},
  cell{3}{5} = {c},
  hline{1-2,4} = {-}{},
}
Privileged country  & Disparate impact          & Bias--d     & Statistical parity          & Bias--s     \\
\textbf{Australia} & \textbf{1.28571428571429} & \textbf{-} & \textbf{0.1}                & \textbf{-} \\
\textbf{Singapore} & \textbf{1.14285714285714} & \textbf{-} & \textbf{0.0571428571428571} & \textbf{-} 
\end{tblr}
}
\label{Resultados después de aplicar mitigación}
\end{table}

\begin{table}[H]
\centering
\caption{Mitigated ranking vs. biased ranking}
\label{table:new-rankings-ndcg}
\resizebox{\linewidth}{!}{%
\begin{tblr}{
  row{1} = {c},
  cell{2}{3} = {r=2}{},
  hline{1-2,4} = {-}{},
}
Country~              & NDCG in mitigated ranking  & NDCG in biased ranking                       \\
\textbf{Australia} & \textbf{0.87261274232549}  & \textbf{0.9816799804069377} \\
\textbf{Singapore} & \textbf{0.861982574659202} &                                                
\end{tblr}
}
\label{Comparación entre el valor de NDCG}
\end{table}

\section{Discussion} 
\label{Capitulo_5}
Now we discuss our results according to research questions in Section \ref{sec:objetivos-preguntas-investigacion}.

\subsubsection{\textbf{R.Q.1:}}

In this work we demonstrate that using a two decimal digits precision, with NDCG = 0.9790262355941447 the best combination of weights for standard features is \{$Category$ = 0.7, $Sentiment$ = 0.05, $Score$ = 0.03, $RevLen$ = 0.22 \}. This NDCG value is better than the values of previous works, see e.g. (0.95) \cite{ALERTme} and (0.552) \cite{MiningInformativeReviews}. We obtained a higher NDCG value because the weights that were used in previous works were always obtained by means of heuristic methods. In this work we exhaustively searched for the best combination of weights using two decimal digits precision.

The best combination of weights also shows that RevLen feature displaces Sentiment, which is a very important feature in feedback processing. This displacement is especially surprising because Sentiment is at a very low level of importance, just above Score.

Thus, we can order features by means of an operator $>$ grater than, and we obtain the following order between the features:

$$Category > RevLen > Sentiment > Score$$

It is important to highlight that this order where RevLen has an important place arises from an exhaustive search of all possible weights of two decimal digits precision assigned for the weighted function that calculates the ranking.

However, if the features are used in isolation, the RevLen feature alone is not more important than other features. Indeed, the order is:

$$Category > Sentiment > Score > RevLen$$

\subsubsection{\textbf{R.Q.2:}}

In this work we show that using two decimal digits precision, Entropy feature can be more important than other features. We show that using the feature set \{Category, Sentiment, Score, Entropy\} and the weights \{$Category$ = 0.27, $Sentiment$ = 0.18, $Score$ = 0.03, $Entropy$ = 0.52\} a nearly perfect ranking is obtained with NDCG = 0.9816799804069377. This NDCG value improves the best combination of weights with standard features. Furthermore, this combination of weights shows that the Entropy feature displaces Category, which is the most important feature if the standard set of features is used.

From the ranking based on Entropy the following order between the features can be deduced:

$$Entropy > Category > Sentiment > Score$$

Observe that it is not intuitive to remove Category from the set of main features because, according to previous work, category is an important feature that determines the ranking of a review.

In addition, it is more intuitive that Entropy displaces RevLen because RevLen gives just only an intuition of the quantity of information contained in a review, i.e., a short review provides less information than a longer review. In this sense, Entropy is a more precise measure for the information contained in a text. Thus, it is more likely that Entropy displaces RevLen.

According to our results, Entropy goes further and in addition to displace RevLen, which was an expected behavior, Entropy displaces all the other features. This displacement is a new discovery because it suggests that the amount of information in a review is more important than its Category or its Sentiment, and even more, it is easier to calculate.






Observe that the combination of standard features is maximized when Category is Bug Report, Sentiment is negative, Score is low, and RevLen is large. This combination of characteristics coincides with a text that contains a high Entropy, indeed, a Bug Report generally includes a description of an error, a text with negative sentiment and a low score generally includes a complaint that is related to a bug.
Therefore, it is feasible that Entropy becomes a feature that summarizes the main characteristic of a review.

\subsubsection{\textbf{R.Q.3:}}

This work shows a computational limit when we increase digits precision. 

According to Table \ref{limite-computacional} the pipeline execution time increases as the decimal digits precision increase. Furthermore, Table \ref{limite-computacional} denotes that there is a limitation in using three digits precision (0.001) since there are 1004006004001 possible combinations of weights, and the pipeline execution time is estimated to be approximately of 52 days.

In addition, Table \ref{limite-computacional} shows that the disk space also increases as the decimal digits precision increase. In this table you can see that there is a limit when using a three digit precision because the disk space used is estimated to be 5069.12GB.

As far as we know, there is no an exhaustive search of the best parameters for a weighted function in previous literature. Weights used in previous works were always obtained by means of heuristic methods because of the combinatorial explosion of weights which pushes computers to their limits. Therefore, previous works are not supposed to show the best weight combination. This work proposes a combination of weights that is the best. In order to achieve our results, it was determined a two decimal digits precision due to the current computing resources.

Thus, we can ensure that our result is the best within a model that uses only two decimals, we cannot ensure anything if we use more decimals.
If we want more decimals, we have to use heuristic methods.







\subsubsection{\textbf{R.Q.4:}}

In this work the NDCG metric decreased when applying bias mitigation. For example, Table \ref{Comparación entre el valor de NDCG} denotes that if bias mitigation is applied, the NDCG decreases to 0.87261274232549 in the case of Australia.

In addition, users from Hong Kong and India are the ones with more reviews of type \emph{Bug} and at the same time \emph{Bug} is the category that has the heaviest weight with respect to \emph{ Feature request} and \emph{Other}. Therefore, this could be the cause of the of bias in favor of Hong Kong and Australia in the dataset that uses the ranking annotated by experts.

Figure \ref{fig:entropy-country} denotes that the user reviews with more Entropy come from Australia. Therefore, this could be the cause of the appearance of bias in favor of Australia in the dataset that uses the best ranking.

Finally, it can be seen that the bias in favor of Singapore is maintained whether the ranking annotated by experts or the best ranking that uses the Entropy feature is used.

\section{Conclusions} 
We report on an exploratory study analyzing user feedback ranking using the technique of weighted function. We found that standard features like category, sentiment and text length can be successfully replaced by Shannon entropy. To this end we performed an exhaustive search of the best weight combination using a precision of two digits. In addition, we show that increasing the precision of weight calculations to a precision of three digits leads to computational limitations which can only be addressed by means of heuristic methods.

While our results need a larger sample size, they hint about the importance of replacing standard attributes which are expensive to calculate, indeed, standard attributes like category and sentiment require machine learning techniques and annotated corpus to train the algorithms. Shannon entropy instead is just a set of few arithmetical operations. 

On the other hand, we show that bias in rankings remains a challenge, particularly across different countries. Moreover, there exists a quality attribute trade--off if we want both ranking precision and fairness. Indeed, we show that applying bias mitigation our levels of nearly optimal NDCG significantly fall.  

We hope that our work motivates research in this field. We encourage researchers to analyze heuristic methods in order to obtain better weight values for a weighted function. We also encourage to discover other possible features like entropy in user feedback processing.

\bibliographystyle{splncs04}
\bibliography{CACIC-rojas-mareco}

\end{document}